\documentstyle[epsfig,12pt]{article}
\begin{document}
\newcommand{\be}{\begin{equation}}
\newcommand{\ee}{\end{equation}}
\newcommand{\bc}{\begin{center}}
\newcommand{\ec}{\end{center}}
\newcommand{\bi}{\begin{itemize}}
\newcommand{\ei}{\end{itemize}}
\newcommand{\bo}{\={o} }
\newcommand{\th}{$\Theta$}

%\title{Evidence of glassy behavior in the dynamics of a homopolymer
%chain from molecular dynamics simulations}
\title{Glassy behavior of a homopolymer from molecular dynamics
simulations}

\author{Nikolay V. Dokholyan$^{1,2}$, Estelle Pitard$^3$,\\
 Sergey
V. Buldyrev$^1$, and H. Eugene Stanley$^1$}

\date{$^1$Center for Polymer Studies, Physics Department,
Boston University, Boston, MA 02215, USA\\
$^2$Department of Chemistry and Chemical Biology, 
Harvard University, 12 Oxford Street, Cambridge, MA 02138, USA\\
$^3$Laboratoire de Physique Th\'eorique et Math\'ematique,
Universit\'e Montpellier II, 34000 Montpellier, France}
%\date{\today}
\maketitle

\begin{abstract}
We study at- and out-of-equilibrium dynamics of a single homopolymer
chain at low temperature using molecular dynamics simulations. The main quantities
of interest are the average root mean square displacement of the
monomers below the theta point, and the structure factor, as a
function of time.  The observation of these quantities show a close
resemblance to those measured in structural glasses and suggest that
the polymer chain in its low temperature phase is in a glassy phase,
with its dynamics dominated by traps. In equilibrium, at low
temperature, we observe the trapping of the monomers and a slowing
down of the overall motion of the polymer as well as non-exponential
relaxation of the structure factor. In out-of-equilibrium, at low
temperatures, we compute the two-time quantities and observe breaking
of ergodicity in a range of waiting times, with the onset of aging.
\end{abstract}

%FACTEUR DE STRUCTURE - DEPENDENCE EN Q
%ANALYSE STATIQUE: EN FONCTION DE LA TEMPERATURE
%RESULTATS: EXPOSANTS
%COMPARAISON AVEC LE MC DES VERRES (REVUE KOB, SCIORTINO, BASCHNAGEL)
%DISCUSSION: PROCHE DES MELTS DE BASCHNAGEL
%            COMPATIBILITE AVEC LA THEORIE MCA, PUISSANCE OU STRETCHED
%            PRESENCE DU VIEILLISSEMENT (QUALITATIF)

%\section{Introduction}

The problem of the dynamics of a single polymer chain has been
addressed in many ways, the simplest theory being the Rouse model
\cite{deGennes}. When the polymer is not a Gaussian chain, but evolves
either to a swollen or collapsed state after a change of external
conditions, the dynamics can also be studied using more or less
elaborate scenarios, either phenomenologically (like the ``sausage''
or ``necklace'' pictures of de Gennes and others \cite{PGG,Goldbart}),
numerically \cite{numerics} or analytically \cite{Orland,Dawson}.
Phenomena similar to polymer collapse can be commonly observed as a
protein folds into its compact native state structure, or unfolds to
an expanded coil (though the protein problem is much more complex than
the homopolymer case); more generally a polymer collapses when it is
put in a poor solvent, and expands in a good solvent. However, testing
the theoretical results with experiments has, thus far, proven
difficult, and one has still little insight as to what happens at a
microscopic level during such dynamics.  It is also expected that the
dynamics at low temperatures is slowed down by traps created by the
entanglements of the chain, and the compactness of the structure in
the collapsed phase. We address this issue here for the first time and
show that these effects can be measured through the one-time and
two-time dependence of the structure factor; we prove that aging is an
intrinsic feature of the dynamics of a long collapsed polymer chain.

We perform the kinetic study of the collapse of a single homopolymer
chain after a temperature quench, obtained using molecular dynamics
simulations. We characterize the different steps that occur during
such collapse.  We show how these steps can be interpreted on a
microscopic level, especially, how to account for the fact that for an
intermediate time regime the dynamics slows down drastically in a way
reminiscent of glassy dynamics. We present results for one- and
two-time quantities and show that, at low temperatures, the system
behaves in an out-of-equilibrium way. Finally, we demonstrate the
similarities of a single homopolymer chain with numerical results
obtained on polymeric systems, such as polymer melts
\cite{Baschnagel}, or glass-formers \cite{Sciortino96,Mossa00}, with
the glassy dynamics observed in a one-dimensional model of homopolymer
collapse \cite{Bouchaud}, or in 3D heteropolymer and homopolymer
dynamics studied via analytical techniques \cite{ModeCoupling}.

%\section{The model}

We study a ``beads on a string'' model of a homopolymer in 3D,
consisting of $N=256$ monomers.  We model the monomers as hard spheres
of unit mass with a square-well attractive potential between every
pair of contacts. The total potential energy of the polymer is
\be
%$
{\cal E} = \frac{1}{2}\sum_{i\neq j=1}^{N} U_{ij} + 
\sum_{i=1}^{N-1}V_{i\, i+1}\, , \label{eq:U}
%$
\ee
where
\be
%$$
U_{ij} = \left\{ \begin{array}{ll}
                   +\infty,   & |{\bf r}_i-{\bf r}_j|\le a_0\\
                   -\epsilon, &  a_0 < |{\bf r}_i-{\bf r}_j|\le a_1\\  
                   0, &  |{\bf r}_i-{\bf r}_j| > a_1\, ,
                  \end{array} 
\right.
%$$
\label{eq:Uij}
\ee
is the matrix of interactions between monomers $i$ and $j$, and
\be
%$$
V_{i\, i+1} = \left\{ \begin{array}{ll}
            0, &  a_0 < |{\bf r}_i-{\bf r}_{i+1}| < a_1\\
            +\infty, &  |{\bf r}_i-{\bf r}_{i+1}|\le a_0, \mbox{ or } 
            |{\bf r}_i-{\bf r}_{i+1}|\ge a_1\,   
            \end{array}\right.
%$$
\label{eq:Vnn}
\ee
is the potential energy of covalent bonds. $a_0/2$ is the radius of
the hard sphere, and $a_1/2$ is the radius of the attractive sphere
($\epsilon=1$, $a_0=8.00$ and $a_1 = 11.28$). The $\theta$-point
temperature for the homopolymer studied is $T_{\theta} \approx 1.75$.
The average radius of gyration of the collapsed globule at $T=0.2 \ll
T_{\theta}$ is $R_G \approx 24.7$.

We employ the discrete molecular dynamics algorithm \cite{Grosberg97}.
To control the temperature of the homopolymer, we introduce a heat
bath of 2746 particles, which do not interact with the homopolymer or
with each other in any way but via collisions \cite{Dokholyan98}. By
changing the kinetic energy of heat bath particles we control the
temperature of the environment. The heat bath particles are hard
spheres of the same radii as the monomers and have unit mass. The
temperature units are in $\epsilon/k_B$. The time unit (see
\cite{Dokholyan98} for details) is estimated from the shortest time
between two consequent collisions in the system between any two
particles.

Next, we compute: {\bf (A)} the root mean square displacement
(rmsd) $R(t_w,t_w+\tau)$ of the polymer between the two times $t_w$
and $t_w + \tau$:
\be
%$$
R(t_w,t_w+\tau) = \left\langle \left[\frac{1}{N} \sum_{i=1}^{N}
\left({\bf r}_i(t_w) - {\bf r}_i(t_w+\tau)\right)^2\right]^{1/2}
\right\rangle \, ,
%$$
\label{eq:sigma}
\ee
where ${\bf r}_i(t_w)$ and ${\bf r}_i(t_w+\tau)$ are the coordinates
of the monomers at times $t_w$ and $t_w +\tau$. The coordinates at
times $t_w$ and $t_w +\tau$ are adjusted with respect to each other so
that: {\it (i)} homopolymer's centers of mass overlap at these two
times; {\it (ii)} their relative orientation is modified using Kabsch
transformation \cite{Dokholyan98}. {\bf (B)} The dynamic structure
factor is defined as
\be
F_q(t_w, t_w+\tau) \equiv 1/N \left< \rho^*_q(t_w) \rho_q(t_w+\tau)
\right>\, ,
\ee
where $\rho_q(t) \equiv \sum_i \exp(i {\bf q \cdot r}_i(t))$. We
compute this quantity for $q=q_{max}$ which corresponds to the maximum
of the equilibrium structure factor and is equal in our system to
$q_{max} = 12/90$. Thermal averaging, $\langle\dots\rangle$, is done
over 10 initial configurations.

%\section{Results}

%\subsection{1-time quantities}

\begin{figure}[hbt]
\centerline{ \vbox{ \hbox{\epsfxsize=8.0cm
\epsfbox{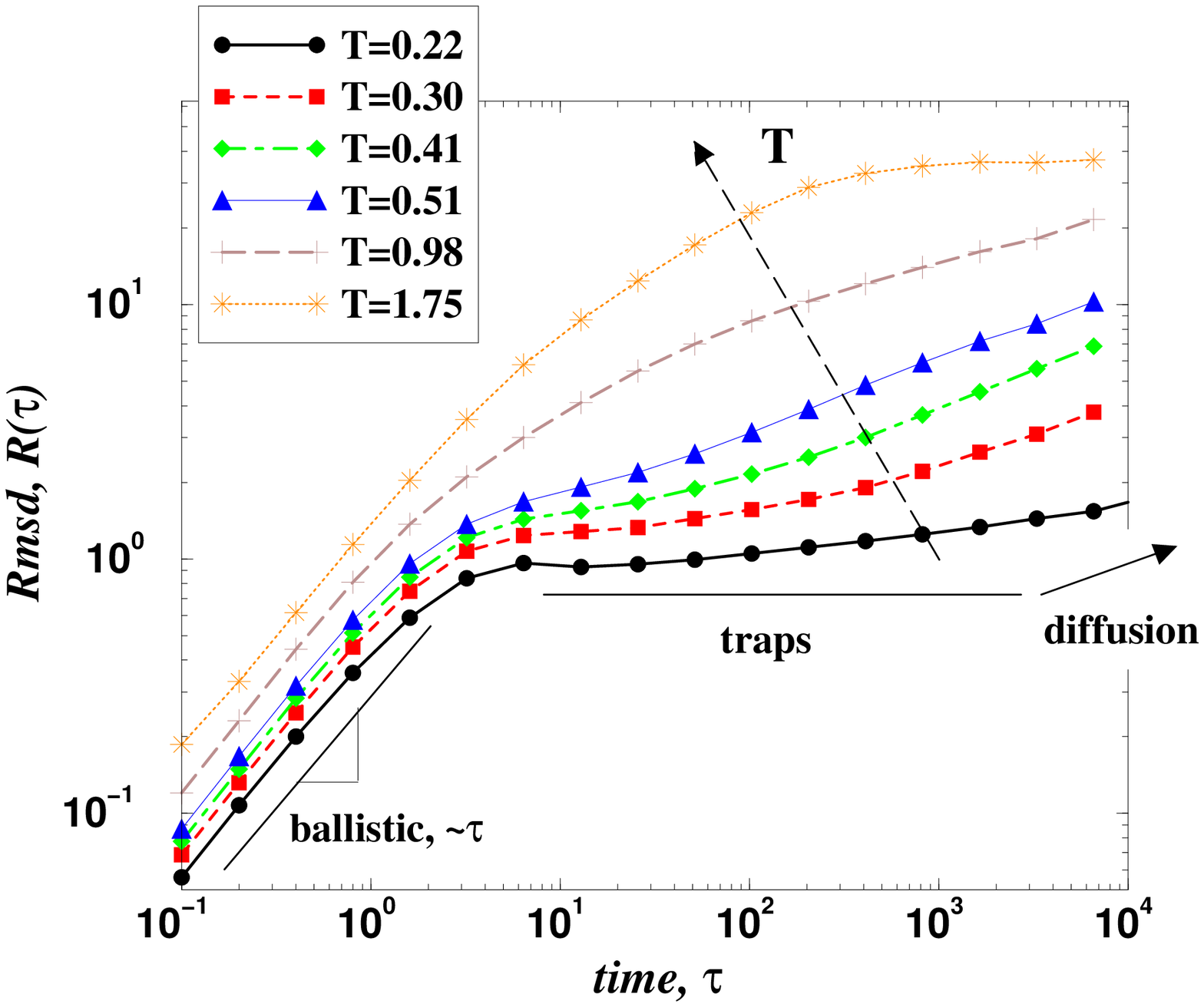}  }}}
\vspace*{0.5cm}
\caption{The dependence of the rmsd $R(\tau)$ on $\tau$ in equilibrium
simulations at various fixed temperatures for a polymer chain of $256$
monomers. Three regimes can be observed: {\it (i)} ballistic (at small
time scales), {\it (ii)} dominated by traps (at intermediate time
scales), and {\it (iii)} diffusive (at long time scales). As the
temperature increases, the second regime becomes less apparent and, at
temperatures around $T^* \approx 0.7$, it disappears.}
\label{fig:1}
\end{figure}

First, we study the polymer at equilibrium, which is achieved by
simulating the polymer at fixed target temperature for approximately
$10^6$ time units. In this case, we expect time-translational
invariance for $R(t_w,t_w+\tau) = R(\tau)$ and $F_q(t_w,t_w+\tau) =
F_q(\tau)$. We simulate the polymer at various temperatures ranging
from 0.2 to 1.75. We find that there are three distinct regions for
the $R(\tau)$ dependence on $\tau$ at low temperatures
(Fig.~\ref{fig:1}). The first region can be viewed as a ballistic
motion of the monomers until they reach their nearest
neighbors. Indeed, we find that the slope of $R(\tau)$ vs $\tau$ on
log-log scale is close to 1 in this region.  Starting at around
$\tau\sim 10$, the rmsd dependence on $\tau$ becomes slower. The lower
the temperature, the flatter is the plot of $R(\tau)$ vs $\tau$ in
this region. Following the dynamics of a single monomer we can explain
the slowing down of $R(\tau)$ by the presence of neighbors which trap
this monomer in a ``cage''. To escape the trap, a collective effort of
the monomer and its neighbors is necessary.  Thus, this region can be
seen as a regime of activated processes that drive monomers beyond
their cages.  At later times the dynamics of the system becomes
diffusive, where we expect and find $R(\tau)\sim \sqrt\tau$, except
for $T=0.2$ where equilibrium could not be achieved during the
simulation time. With the increase of temperature the trapping time
becomes smaller and we observe a change in the curvature of the plot
$R(\tau)$ vs $\tau$, where the trapping plateau actually
disappears. The change of curvature occurs at a temperature around
$T^* \approx 0.7$.

\begin{figure}[hbt]
\centerline{ \vbox{ \hbox{\epsfxsize=8.0cm
\epsfbox{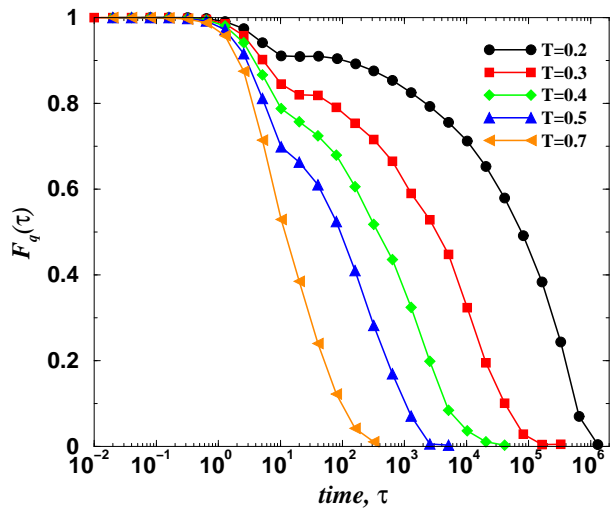}  }}}
\vspace*{0.5cm}
\caption{The dependence of the structure function, $F_q(\tau)$ on
$\tau$ in equilibrium simulations at various fixed temperatures. We
choose $q=q_{max}$, which corresponds to the nearest neighbor shell of
the polymer.}
\label{fig:2}
\end{figure}

In order to compare the dynamic behavior of our homopolymer with that
of known glassy systems (see e.g. \cite{Sciortino96,Mossa00}), we
study the structure factor $F_q(\tau)$ at equilibrium for various
temperatures (Fig.~\ref{fig:2}). The mode coupling theory (MCT) for
glasses \cite{Goetze}, has shown that in such systems, the scaling of
the structure factor behaves non-exponentially with time and predicts
power-law or stretched-exponential dependence according to the time
regime. We find that $F_q(\tau)$ has the properties predicted by MCT
at low temperatures (see Fig.~\ref{fig:2}): {\it (i)} initially
$F_q(\tau)$ decays as a power-law function $1 - F_q(\tau)/ F_q(0) \sim
\tau^{-g}$ to a ``plateau'', $F_q^c$; {\it (ii)} $F_q(\tau)$
approaches the plateau, $F_q^c$, as a power-law function ($F_q(\tau) -
F_q^c \sim (\tau/ \tau_0)^{-a}$); {\it (iii)} $F_q(\tau)$ departs from
the plateau as a power-law, $F_q^c - F_q(\tau) \sim (\tau/ \tau^*
)^{b}$ ($\beta$-{\it relaxation}); {\it (iv)} in a larger time sale
($\alpha$-{\it relaxation}) $F_q(\tau)$ decays to 0 as a stretched
exponential, $F_q(\tau) \sim \exp \left( - (\tau / \tau^*)^{\beta}
\right)$.

We find that {\it (i)} the exponent $g = 1.9 \pm 0.2$, which is in
agreement with the expected exponent $g = 2$ due to the ballistic
motion of the particles at the time scales equal to the size of a
typical cage divided by the typical velocity of a particle; {\it (ii)}
exponent $a$, $0.3<a<0.7$, is difficult to determine due to
uncertainty in $F_q^c/F_q(0) \approx 0.8 - 0.9$; {\it (iii)} the
departure from the plateau, $F_q^c \approx 0.9$, is well fitted by a
power-law with the exponent $b = 0.3 \pm 0.02$, which is almost
independent of $T$; {\it (iv)} the $\alpha$-relaxation is well fitted
by a stretched exponential with the exponent $\beta = 0.52\pm 0.03$,
which is also independent of $T$ \cite{footnote1}.

%\subsection{2-time quantities}

\begin{figure}[hbt]
\centerline{ \vbox{ \hbox{\epsfxsize=8.0cm
\epsfbox{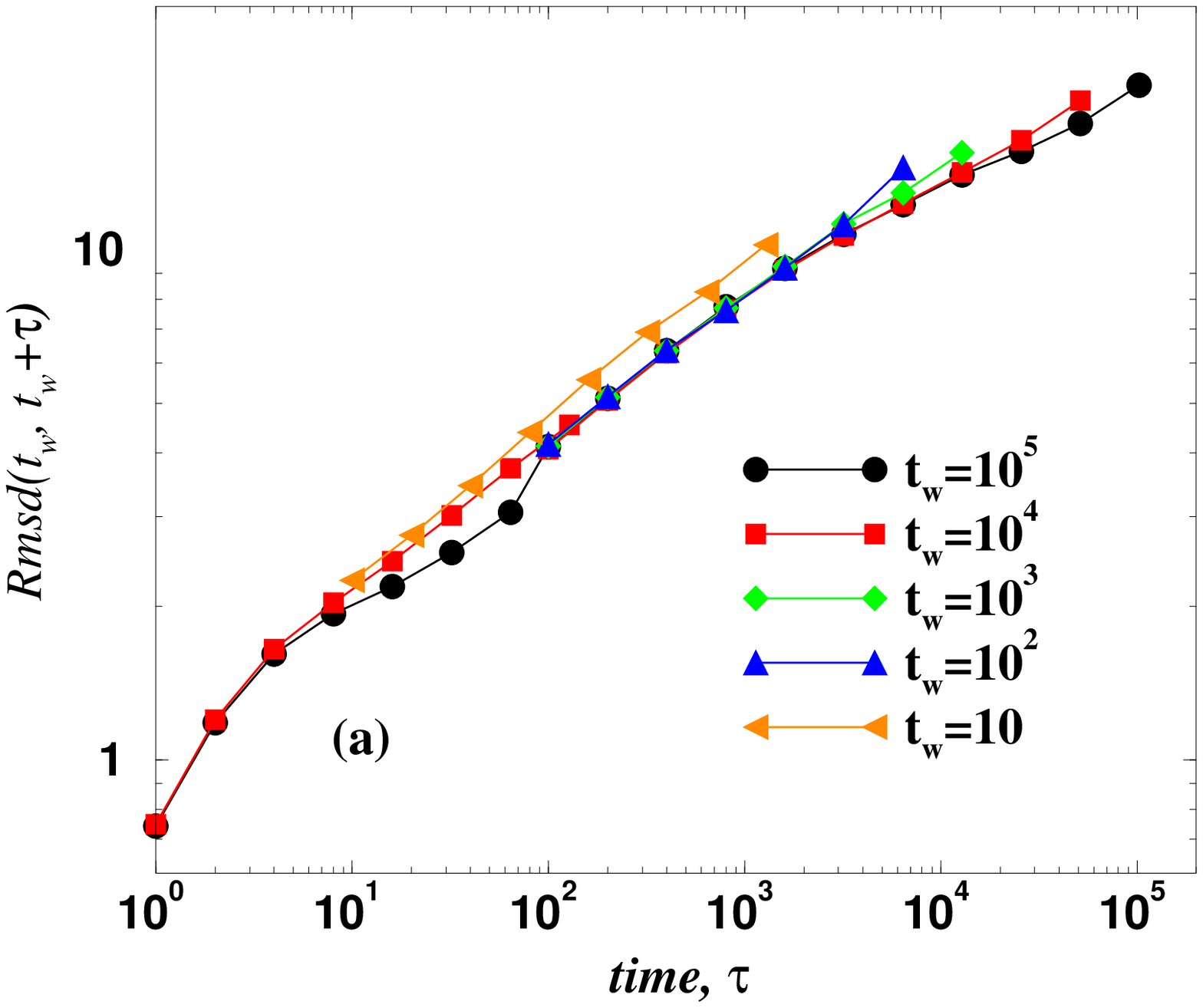}  }}}
\vspace*{0.5cm}
\centerline{ \vbox{ \hbox{\epsfxsize=8.0cm
\epsfbox{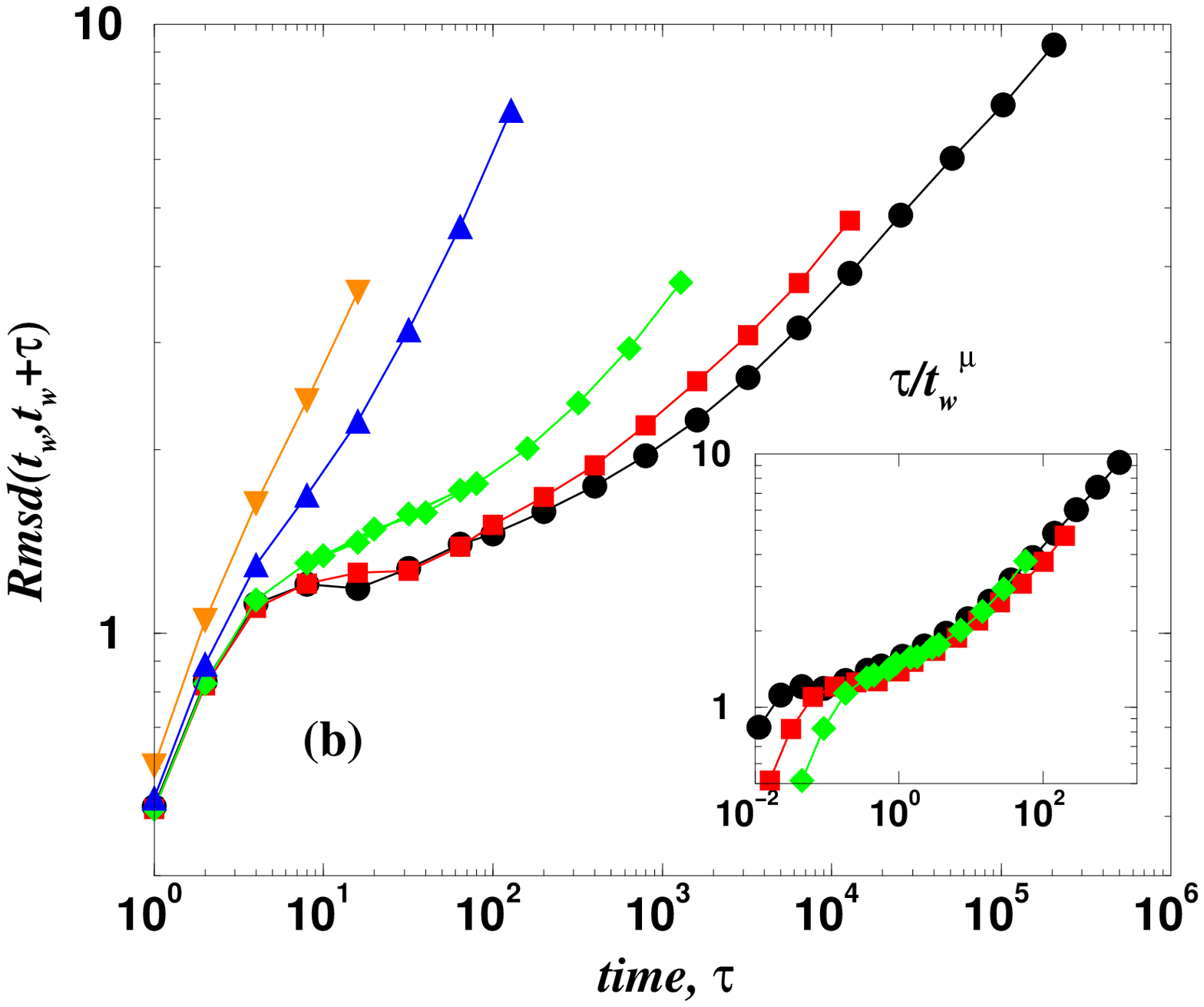}  }}}
\vspace*{0.5cm}
\caption{ The dependence of the rmsd $R(t_w, t_w+\tau)$ on $\tau$ in
quenching simulations from $T_{\theta} \approx 1.75$ to (a) $T_t=0.6$
(no aging observed) and (b) $T_t=0.3$ (aging) for various waiting
times, $t_w$. Inset: rescaling of the curves for $t_w=10^5, 10^4,
10^3$, with respect to $\tau/t_w^{\mu}$, where $\mu \simeq 0.45$.  The
drastic difference between (a) and (b) points towards the existence of
a temperature at which the polymer dynamics becomes non-ergodic.}
\label{fig:3}
\end{figure}

\begin{figure}[hbt]
\centerline{ \vbox{ \hbox{\epsfxsize=8.0cm
\epsfbox{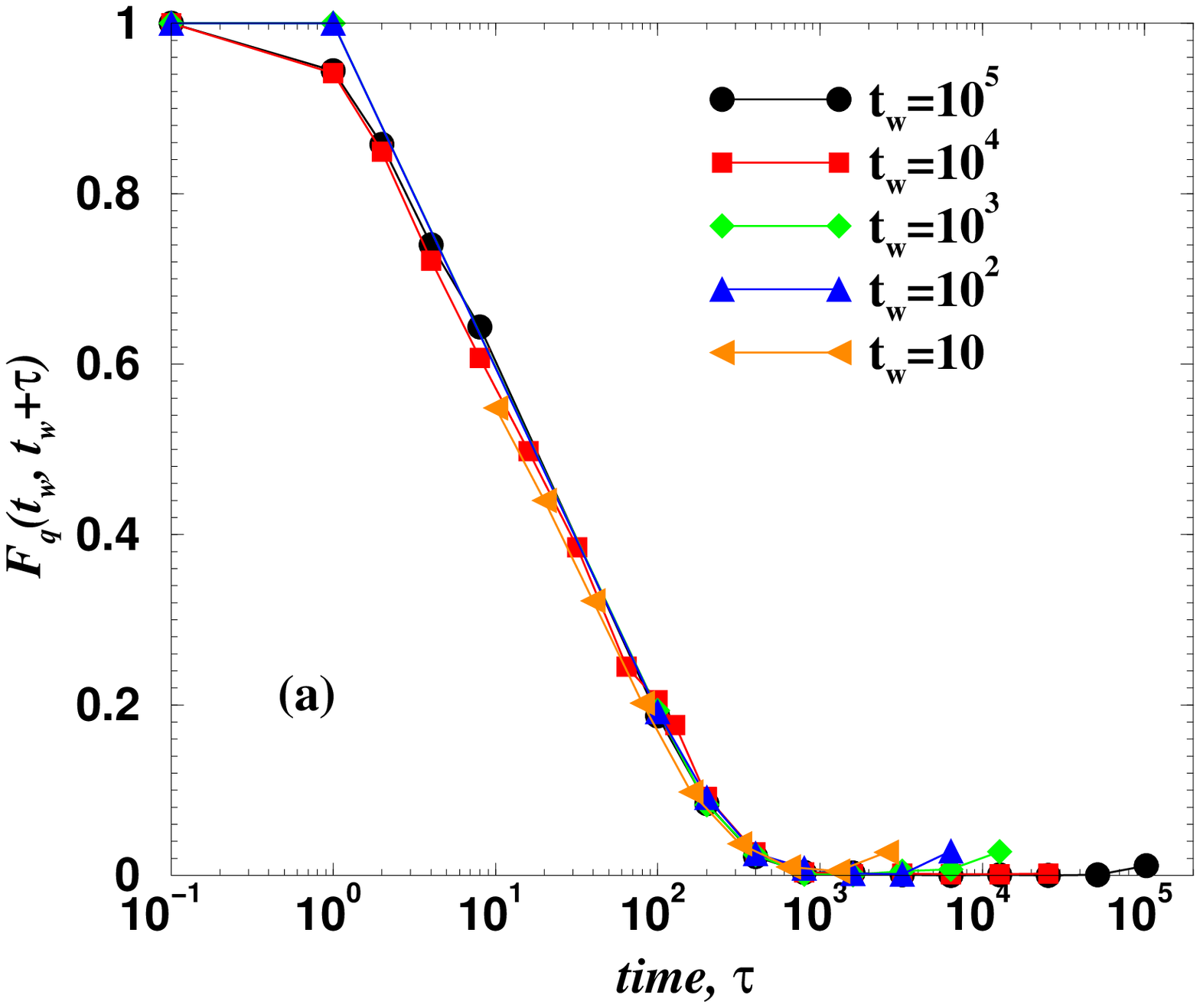}  }}}
\vspace*{0.5cm}
\centerline{ \vbox{ \hbox{\epsfxsize=8.0cm
\epsfbox{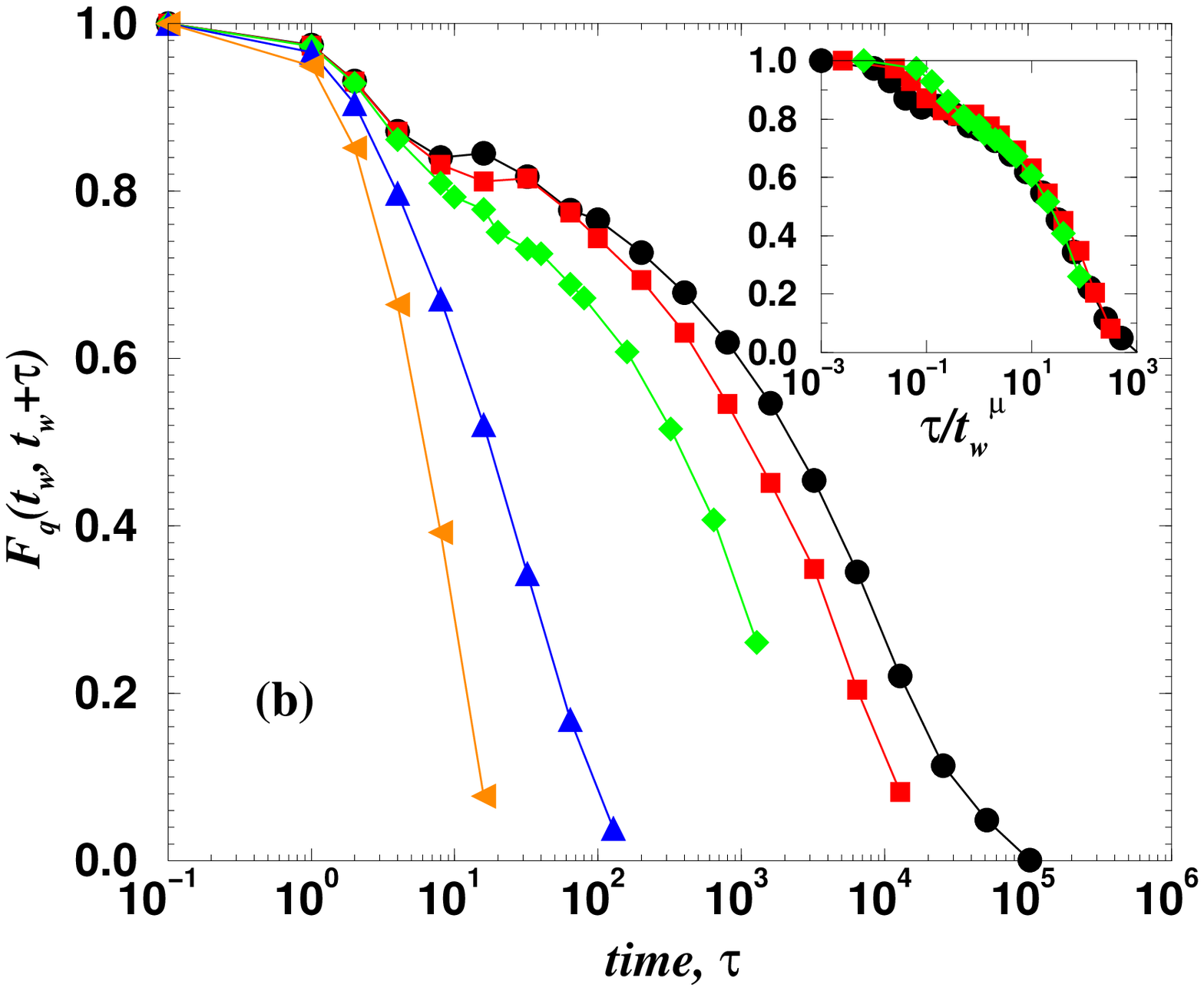}  }}}
\vspace*{0.5cm}
\caption{
The same as Fig.~\protect\ref{fig:3} but for the structure factor
$F_q(t_w, t_w+\tau)$. The exponent $\mu$ is equal to $0.45$.}
\label{fig:4}
\end{figure}

The sign that a system is of out-of-equilibrium is the violation of
time-translational invariance of quantities such as $R(t_w,t_w+\tau)$
and $F_q(t_w,t_w+\tau)$, called aging \cite{Spinglasses}. To recover
aging phenomena, we study the dependence of $R(t_w,t_w+\tau)$ and
$F_q(t_w,t_w+\tau)$ on $t_w$. We quench the polymer from a high
temperature ($T=T_{\theta}$) to lower target temperatures $T_t$ and
then let the polymer evolve at $T_t$. We start computing the two-time
quantities after some waiting time $t_w$ after the quench
(Figs.~\ref{fig:3} and ~\ref{fig:4}).

We find that $R(t_w,t_w+\tau)$ and $F_q(t_w,t_w+\tau)$ do not exhibit
dependence on $t_w$ (see Figs.~\ref{fig:3}a and ~\ref{fig:4}a) if we
quench the polymer to temperatures above $T_t = T_b\approx 0.6$. At
$T_b=T_t\approx 0.6$, $R(t_w,t_w+\tau)$ and $F_q(t_w,t_w+\tau)$ become
dependent on $t_w$, indicating a slowing down of the dynamics as $t_w$
increases. Thus, $T_b$ is the temperature at which we observe the
breaking of the time translational invariance. At $T_t = 0.3$ (see
Figs.~\ref{fig:3}b and \ref{fig:4}b) we observe a separation of the
$R(t_w,t_w+\tau)$ and $F_q(t_w,t_w+\tau)$ vs $\tau$ curves with
respect to waiting time $t_w$.  In the large $t_w$ and $\tau$ regime,
we show that both $R(t_w,t_w+\tau)$ and $F_q(t_w,t_w+\tau)$ can be
rescaled as a function of $\tau/t_w^{\mu}$ (see Figs ~\ref{fig:3}b and
~\ref{fig:4}b). The exponent $\mu$ is equal to $0.45$ for $T_t=0.3$
and is the same for $R(t_w,t_w+\tau)$ and $F_q(t_w,t_w+\tau)$. $\mu$
depends strongly on the temperature and it increases as the temperature
decreases.

In a compact state (at $T_t\approx 0.3$), the polymer exhibits glassy
behavior on the time scale accessible to our simulations. The spatial
rearrangements of monomers are slower due to the existence of
energetic barriers. Due to the finite length of our homopolymer, we
expect that at long times, equilibrium is achieved and the
time-translational invariance is restored.

%\section{Discussion}
We find that the dynamics of a polymer chain at low temperatures is
similar to the dynamics of strongly frustrated systems, such as
glasses \cite{Kob}, and of strongly disordered systems, such as
spin-glasses \cite{Spinglasses}. Equilibrium plots similar to
Fig.~\ref{fig:1} and Fig.~\ref{fig:2} have been previously found in
studies of polymer melts \cite{Baschnagel}.  Here we also demonstrate
that the existence of aging is an intrinsic feature of the dynamics of
a polymer chain at low temperatures.  We expect the microscopic
mechanisms responsible for the slowing down of the dynamics to be
similar to those found in glasses, and they can be easily identified
in our case as geometrical frustration.  The two temperatures, $T^*
\approx 0.7$ and $T_b\approx 0.6$, at which trapping starts to
dominate the dynamics and time translational invariance is broken, are
close to each other and may well correspond to a unique dynamical
transition temperature.  The analysis of the equilibrium data is
consistent with some of the results predicted by the MCT theory. From
the point of view of aging, we do not know any theory for 3D polymer
dynamics that would predict a $\tau/t_w^{\mu}$ scaling. However, this
scaling (with $\mu < 1$) can be found in a number of systems,
including real spin-glasses \cite{Spinglasses}, and in models of traps
\cite{StAndrews} or in a model of a one-dimensional polymer
\cite{Bouchaud}; whereas a $\tau/t_w$ scaling is more specific to
near-mean-field models \cite{Spinglasses,ModeCoupling}.

A single homopolymer provides a new example of a non-disordered system
where geometry alone induces slow relaxation and aging. A number of
polymeric systems in a compact state (single polymeric molecules,
polymer melts, collapsed proteins) share the dynamical features
described here. Some open questions remain, such as the very long time
regime of the relaxation which is hard to characterize due to long
computational times.

%\section{Acknowledgments}

We thank A. Coniglio, W. Kob, S. Mossa, F. Sciortino, and E. I.
Shakhnovich for helpful discussions. NVD is supported by NIH
postdoctoral fellowship GM20251-01. EP is supported by NIH grant GM
52126 and by Centre National de la Recherche Scientifique (France).

\begin{thebibliography}{99}
\bibitem{footnote1} {Another prediction of the MCT is that the
relaxation time $\tau^*$ during the $\alpha$-relaxation scales with
the temperature as $\tau^* \simeq \tau_0 |T-T_c|^{-\gamma}$. The
exponent $\gamma$ is not independent but is related to exponents $a$
and $b$ as $\gamma = 1/2a + 1/2b$. We estimate the exponent $\gamma=5
\pm 3$ and $T_c= 0.03 \pm 0.01$ from the $F_q(\tau)$ dependence on
$\tau$ for various $T$.}
\end{thebibliography}
\end{document}